\documentstyle[12pt]{article}
\begin{document}
\begin{center}
\section*{\bf The effect of Planck scale space time fluctuations on
Lorentz invariance at extreme speeds} 
\vspace{1.5mm}

Richard Lieu
 
Department of Physics, University of Alabama, Huntsville,
AL 35899, U.S.A.
\end{center}
 
\vspace{1.5mm}
 
\noindent
{\bf Abstract}
 
\noindent
The starting point of this work is the axiomatic existence 
of a smallest measurable interval, viz. the Planck time
$t_P$, set by quantum fluctuations in the vacuum metric tensor.  By the 
Relativity Principle, the same limit must then apply to the accuracy of all 
clocks which register time of events in their own frames.  Further, it implies
that the ordinary meaning of distance also ceases in the same manner beyond  a
scale $l_P = c t_P$.  We demonstrate that quantum space-time, if real, may 
be made manifest by observing very energetic collisions, defined as 
interactions which occurred with the center-of-mass frame $\Sigma'$ (of the 
participating bodies) moving at a high speed $v$ relative to our laboratory 
frame $\Sigma$.  In such situations, the initial conditions of the interaction 
are determined from direct measurements of the ultra-energetic particles or 
photons by instruments aboard $\Sigma$: they gather a raw dataset ${\cal S}$
which are subject to the limiting uncertainties $\sim (t_P, l_P)$.  Yet a 
meaningful (i.e. experimentally verified) version of the interaction is one
where the interaction is viewed from frame $\Sigma'$.  Since no instruments 
aboard $\Sigma'$ have taken any data, the way we proceeded is by a Lorentz 
transformation of ${\cal S}$ from $\Sigma$ to $\Sigma'$.  Beware however that 
the resulting ${\cal S'}$ no longer consists of raw data, i.e. Lorentz
distortions of probability distributions render the uncertainties non-Planckian 
- in fact, it will be shown that as $v \rightarrow c$ they are $\gg$ 
$(t_P, l_P)$, 
and are no longer negligible.  Examples will given to indicate how a proper
interpretation of the most current high energy cosmic and gamma-ray data 
necessitates incorporation of the said effect.

\newpage

At present no one knows the design of the most precise clock.
There is nonetheless a broad agreement that
the unsurpassable limit is $\sim$
$t_P$ (see, e.g. Wilczek 2001), because to achieve such an
accuracy a large amount of energy
(hence mass) has to be confined
within a small region of space - the result is a
black hole.  We shall therefore
begin with a standard deviation $\sigma_t$, which is $\sim t_P$, and
{\it postulate} that all clocks which keep time aboard their
respective  inertial 
platforms carry the same intrinsic uncertainty of $\sigma_t$
(cf Principle of Relativity).
The measurement of space by an inertial observer
must then be subject to an error 
$\sigma_x = c \sigma_t$, since by special relativity this simply
involves light beams and synchronized clocks.  Scale sizes like
$t_P$ and $l_P$, though extremely small, may soon be probed
by interferometric experiments (Camelia 2001, Abramovici 1996).
Here we propose another way of testing the discreteness of time,
by observing the behavior of ultra-energetic particles and photons.

Let us now consider the effect of frame transformation on space-time
data.  Suppose two inertial observers $\Sigma$ and $\Sigma'$ construct
their own frames of reference having spatial axes oriented in the
same way, and that
$\Sigma'$ moves
relative to $\Sigma$ with a speed $v$ along the $+x$ direction\footnote{
For uniform motion we assume that the uncertainty in $v$ due to space-time
fluctuations may be made arbitrarily small by adopting a large interval
for the measurement of $v$}.  Suppose further that $\Sigma$ and $\Sigma'$
used instruments aboard their own platforms to
locate an event, with the results $(x,y,z,t)$ and
$(x',y',z',t')$ respectively, which we shall henceforth refer to
as {\it raw data} (to distinguish them, in particular, from data
which emerge from frame transformations).
According to our postulate, intrinsically both
sets of coordinates are equally 
inaccurate to within
$\pm c \sigma_t$ for $x$, $y$,
$z$ (and $x'$, $y'$, $z'$) and $\pm \sigma_t$ for $t$ (and $t'$).  
This indicates the absence of a preferred frame, and remains valid
so long as the two observers measure the event by their own
clocks and light beams.  The symmetry will be broken, however, 
if e.g. $\Sigma'$
fails to take his own data, but monitors the event
solely from the data taken
by $\Sigma$.  In that case, the original raw data 
of $\Sigma$
would appear meaningless
to $\Sigma'$ until they are {\it processed} by the Lorentz transformation 
${\cal L}$ which necessarily alters the uncertainties.

For a quantitative estimation of the effect, we note that according to
${\cal L}$:
\begin{equation}
x' = \gamma (x-vt)~~,~~y'=y~~,~~z'=z~~,~~t'=\gamma(t-\frac{vx}{c2}),
\end{equation}
where $\gamma = (1-v2/c2)^{- \frac{1}{2}}$.  Obviously, then,
$y'$ and $z'$ remain uncertain by $c \sigma_t$.  To calculate
the uncertainies of $x'$ and $t'$ one must add those of $x$ and $t$
in quadrature, because these two coordinates were directly 
and separately measured
by $\Sigma$ their errors are {\it independent} (and, for the same reason,
of magnitude $c \sigma_t$ and $\sigma_t$ respectively).  Thus we have:
\begin{equation}
\sigma_{x'}^2 = \gamma2 (\sigma_x2 + v2 \sigma_t2)~=~
c2 \sigma_t2 \frac{1+\frac{v2}{c2}}{1 - \frac{v2}{c2}}~~;~~
\sigma_{t'}^2 = \gamma2 (\sigma_t2 + \frac{v2}{c4} \sigma_x2)~=~
\sigma_t2 \frac{1+\frac{v2}{c2}}{1 - \frac{v2}{c2}}.
\end{equation}
In general, Eq. (2) indicates that $\sigma_{x'} \geq \sigma_x$,
$\sigma_{t'} \geq \sigma_t$.  Further, $\sigma_{x'}$ and $\sigma_{t'}$
are no longer independent errors.  Unlike $\sigma_x$ and $\sigma_t$,
they cannot be added in quadrature.

It should be emphasized that, among the
consequences which may ensue from Eq. (2), violation of the Principle
of Relativity is not one of them.  The coordinates $x'$ and $t'$
carry correlated errors larger than the intrinsic values because they are not
raw data.
Had $\Sigma'$ performed his own measurements, the results
would fundamentally have been limited by the
quantum space-time fluctuations 
$(\sigma_t, c \sigma_t)$ in exactly the same way
as those of $\Sigma$.  Any such data
taken directly by $\Sigma'$, when compared
with the transformed data of $\Sigma$
(which by Eq (2) carry larger errors) could lead to disagreement.
Conversely, if $\Sigma'$ were to transform by ${\cal L}^{-1}$ his own data to
the $\Sigma$ frame, the results would likewise be subject to the same
error amplification and correlation as depicted in Eq (2), and
could disagree with
the measurements done by $\Sigma$ himself.  Moreover, 
the symmetry imposed by the Principle of Relativity tells
us that the two observers
could continue to argue about `who is correct'.  Neither in fact is
on a privileged platform.  Rather, the source of the
discrepancy lies with the fact that when
raw data are processed ${\cal L}$ they inevitably emerge with larger
than intrinsic (Planckian) uncertainties.  This is a feature of ${\cal L}$
which we have hitherto ignored.

The significance of Eq (2) depends on the situation it applies to.
As an example, let
$\Sigma$ measure the decay time
of a fast-moving particle, which we
assume to be at rest relative to $\Sigma'$.  The result, denoted
by $\tau$, is simply
obtained by taking the difference between two
time coordinates, each of which is subject independently to
the same random uncertainty $\sigma_t$, i.e. $\tau = t_2 - t_1$ and
$\sigma_{\tau} = \sqrt{2} \sigma_t$.
Likewise, if $\Sigma'$ were to measure
this particle directly his result $\tau'$ will also be accurate to
$\pm \sqrt{2} \sigma_t$.  Now
suppose
no data were ever gathered aboard
$\Sigma'$, and the only information available
concerning the rest lifetime $\tau'$ of the particle is the
measurement of $\Sigma$ after Lorentz transformation to the
$\Sigma'$ frame.  From
Eq (2) we see that ${\cal L}$ enlarges intrinsic uncertainties by
the factor $\sqrt{2} \gamma$
when $v \rightarrow c$.  This means, had $\Sigma'$ done his
own measurement and compared it with the transformed data of
$\Sigma$ there may, in the case $\gamma \gg 1$, be substantial
discrepancies on the value of the
rest life $\tau'$.  One could insist upon a shorter lifetime than the other,
which renders it inevitable that they could disagree with
each other in a random manner on whether the
decay has taken place or not (the `time sequencing' problem which
will not be discussed in detail here).  The transformed data of
$\Sigma$ carry a relative (percentage) uncertainty 
of $2 \gamma \sigma_t/ \tau'$, which exceeds that
for the raw data of $\Sigma'$ and $\Sigma$ by a factor of $\sqrt{2} \gamma$
and $\sqrt{2} \gamma2$ respectively.  The reliance of $\Sigma'$ on
the measurement of $\Sigma$ 
can no longer be justified when
this relative error $2 \gamma \sigma_t/ \tau'$ is $\geq 1$, which happens for
sufficiently large $\gamma$.

In the present work we pursue to some
depth the impact of the above development
on another common usage of ${\cal L}$.
Often the {\it physics} of
ultra-energetic quanta cannot at all be perceived in our laboratory frame 
$\Sigma$ - we must transform its parameters which we measured aboard $\Sigma$
to another frame $\Sigma'$ before any sensible conclusions
on the nature and fate of the subject can
be drawn.   As an example,
consider
a $10^{20}$ cosmic ray proton in the vicinity of a 3K microwave
background photon.
While we may have data about the two bodies taken by
laboratory detectors working in the $\Sigma$ frame, no experiments 
done aboard $\Sigma$  have ever directly witnessed how they may
interact.  However, from the viewpoint of the proton rest
frame $\Sigma'$ a 3K photon appears, according to our laboratory data
of the two particles after transformation by
${\cal L}$, as having sufficient energy to cross the photo-pion
production threshold.
This reveals a familiar occurence in our own
environment (viz. pions from photon-proton interactions)
which, by the Principle of Relativity, must happen aboard $\Sigma'$
as well.  Thus we conclude that aboard $\Sigma$ one necessarily
encounters pions, and the required energy is now drawn from the
fast proton.
The logic of the above reasoning is sound except for one point:
it presumes that ${\cal L}$ remains highly accurate when
the relative motion between $\Sigma$ and $\Sigma'$ involves a
Lorentz factor as large as $\gamma \sim 10^{11}$.

In order to address this point, we must realize that once time and
space fluctuate intrinsically the components of all four-vectors will
behave similarly.  The following example shows how to 
quantitatively secure the
connection.
Consider a
plane light  wave.  If by some means the wave frequency
$\nu$ is very precisely known in the regime
$\nu > 1/\sigma_t$ we will be able to use the wave as a
`super-clock' to overcome
quantum limitations - which by assumption is impossible.  An equivalent
manner of expression would state that any value of $\nu$
can only be determined to an accuracy $\sigma_{\nu}/\nu \sim \nu \sigma_t$.

A more rigorous approach to the problem, with something similar to
the above as an outcome,
is afforded by the
following abstraction of the measurement
procedure.
Let the wave phase be written as
$\phi = (\omega t - kx cos \theta)/c$ where $\theta$ is
the angle which the wavevector ${\bf k}$ makes w.r.t. $+x$.
All the basic information about the wave have to do with one
cycle of oscillation when:
\begin{eqnarray}
\phi \rightarrow \phi + 2 \pi~~as~~x \rightarrow
x + \Delta x~~and~~t \rightarrow t + \Delta t.
\end{eqnarray}
Denoting any possible variations in
the $\Delta$'s by the $\delta$'s, we have:
\begin{equation}
0 = \delta \phi =\omega \delta t - k cos \theta \delta x
 + \Delta t \delta \omega -
\Delta x \delta (k cos \theta)
\end{equation}
Eq (4)  embodies the manner by which space-time fluctuations lead to
limitations in our knowledge of the wave properties.
To elucidate this, we examine the simplest
case $\theta = 0$, and note that
the {\it angular} frequency is by definition
$\omega = 2 \pi/\Delta t$ where
$\Delta t$ refers to the time {\it interval} 
(cf. the aforementioned $\tau$) satisfying
Eq (3) in the limit of vanishing space interval, $\Delta x = 0$, during
which Eq (4)
reduces to
$\delta \omega/\omega = -\delta t/\Delta t + cos \theta (\delta x/c \Delta t)$.
The limiting $\omega$ uncertainty is obtained, by treating $(\delta t,
\delta x)$ as independent random errors of magnitude $(\sqrt{2} \sigma_t,
\sqrt{2} c \sigma_t)$, to be:
\begin{equation}
\sigma_{\omega} = \frac{\sigma_t \omega2}{\sqrt{2} \pi} (1 + cos2 \theta)^
{\frac{1}{2}},
\end{equation}
or simply $\sigma_{\nu}/\nu \sim \nu \sigma_t$, as before.

Likewise, the quantity $k cos \theta$, defined as $= 2 \pi/\Delta x$
where $\Delta x$ refers to the value satisfying
Eq. (3) in the case $\Delta t =0$.  Thus, using again Eq (4) and
arguments similar to those which follow  this equation lead to the result:
\begin{equation}
\sigma_{k cos \theta} = \frac{\sigma_t \omega2 |cos \theta |}{\sqrt{2} \pi c}
(1 + cos2 \theta)^{\frac{1}{2}}
\end{equation}
where in deriving Eq (6) it is also assumed that a real (on-shell) photon
satisfying $\omega/k = c$ is detected in our laboratory.

Once the original data of $\Sigma$ are Lorentz transformed to the $\Sigma'$
frame the frequency will assume a value $\omega'$ given by:
\begin{equation}
\omega' = \gamma [\omega - v (k cos \theta)]
\end{equation}
From the foregoing treatment it is plain that our
laboratory `energy-momentum'  measurements
of $\omega$ and $(k cos \theta)$, like those of $t$ and $x$, must be
regarded as independent.  Thus when $\omega'$ is inferred from $\omega$
and $k cos \theta$ by means of Eq (7), the result carries an error
$\sigma_{\omega'}$ of magnitude:
\begin{equation}
\sigma_{\omega'} = \gamma (\sigma2_{\omega} + v2 \sigma2_{k cos \theta})
^{\frac{1}{2}} =
\frac{\gamma \omega2 \sigma_t}{\sqrt{2} \pi}
(1 + cos2 \theta)^{\frac{1}{2}} \left[1 + \frac{v2}{c2} cos2 \theta
(1 + cos2 \theta) \right]^{\frac{1}{2}},
\end{equation}
where the final form of Eq (8) was obtained with the help of Eqs (5) and (6).
Then it follows from Eq (7) that:
\begin{equation}
\frac{\sigma_{\omega'}}{\omega'} = \frac{\omega \sigma_t}{\sqrt{2} \pi}
\left(1 - \frac{v cos \theta}{c} \right)^{-1}
(1 + cos2 \theta)^{\frac{1}{2}}
\left[1 + \frac{v2}{c2} cos2 \theta (1 + cos2 \theta) \right]^{\frac{1}{2}}
\end{equation}

We now apply our results to
the special case of $\theta = 0$ ($\Sigma'$ moving relative to
$\Sigma$ in the same direction as the photon wavevector) 
and $v \rightarrow c$ ($\gamma \gg 1$) Eqs (7) and (9) simplify to:
\begin{equation}
\omega' = \frac{\omega}{2 \gamma}~~{\rm and}~~
\frac{\sigma_{\omega'}}{\omega'} = \frac{2 \sqrt{3}}{\pi} 
\gamma2 \omega \sigma_t
\end{equation}
As can be seen in Eq (10), there is an accuracy deterioration when
the measurement of $\Sigma$ is referred to
a frame $\Sigma'$ where the frequency $\omega'$ is $\ll \omega$.  More
precisely,
upon comparison of Eq (10) with Eq (5) it is apparent that
$\sigma_{\omega'}/\omega'$ exceeds
the relative error $\sigma_{\omega}/\omega$ intrinsic to
raw data of $\Sigma$ by a factor $2 \sqrt{3} \gamma2$.  It
is also important to assess the ultimate quality
of any measurement directly performed aboard $\Sigma'$.
Had such been available, one would have repeated the reasoning
which led to Eq (5), with $(\omega', \theta'= 0)$ now replacing
$(\omega, \theta = 0)$, to conclude upon
a relative error
$\sim \omega' \sigma_t/\pi = \omega \sigma_t/(2 \pi \gamma)$,
which is less than that of the transformed data of $\Sigma$, Eq (10),
by the factor $4 \sqrt{3} \gamma3$.  This means {\it what we believe
as the frequency seen by observer $\Sigma'$ could differ very
substantially from the actual value witnessed by that observer}.
In fact, Eq (10) indicates that 
the point at which we cannot meaningfully predict $\omega'$
using our laboratory measurement of $\omega$
occurs when:
\begin{equation}
\frac{\sigma_{\omega'}}{\omega'} \sim \gamma2 \omega \sigma_t \geq 1.
\end{equation}
A similar calculation reveals that in the limit of
Eq (11) the `momentum'
parameter $k' cos \theta'$ will also undergo large excursions (yet correlated
with those of $\omega'$).

Further, the argument can be repeated, with similar
conclusions, to a situation where the energy of
an ultra-relativistic particle is measured
in the laboratory frame $\Sigma$ and transformed to 
the particle rest frame $\Sigma'$.  For the purpose of calculating intrinsic
uncertainties, the
laboratory 4-momentum 
$(E,{\bf p})$ of such a particle may be treated in the same way as that of
the photon $\hbar (\omega, {\bf k})$, with an on-shell
relation\footnote{The
relative error in the $E \approx pc$ approximation is $O(1/\gamma2)$
for $\gamma \gg 1$.  As long as this is $\ll$ the relative error in $E$
caused by space-time fluctuations, viz. $\sigma_E/E \sim
E \sigma_t/\hbar$ (which is the case in our example of a
10$^{20}$ eV proton) the approximation is valid.}
of $E = (p^2c2 +m^2c4)^{\frac{1}{2}} \approx pc (\gg mc2)$.
Then the transformed energy $E'$
carries a relative uncertainty $\sigma_{E'}/E'$ which likewise
exceeds that of the original measurement of $\Sigma$, viz. $\sigma_{E}/E$,
by $2 \sqrt{3} \gamma2$ times.  Again, analogous to Eq (11), our own
data become irrelevant towards predicting actual rest frame
experiences when:
\begin{equation}
\frac{\sigma_{E'}}{E'} \sim \frac{\gamma2 E \sigma_t}{\hbar} \geq 1.
\end{equation}
When the limit given by Eq (12) is reached the particle energy
$E'$ fluctuates substantially about $mc2$.  Here and beyond, a key
identifying characteristic of the particle, viz. its rest mass,
ceases to be meaningful, nor can one assign a special status to
the rest frame.

We now consider two applications.  In each
case presented, we point out the interesting phenomenon of an observational
anomaly occurring at a parameter regime where the effect discussed
in this paper is non-negigible.  We then provide a qualitative 
yet compelling
argument to explain
why the anomaly may be a manifestation of  quantum space-time.
To proceed, a value for the basic
parameter $\sigma_t$ is necessary.  As discussed earlier, it makes
sense to adopt the Planck time $t_P \approx 5.4 \times 10^{-44}$ s as
a conservative estimate.  

Our first example concerns 
the laboratory frame detection of $\sim$ 25 TeV 
($\omega \sim 4 \times 10^{28}$ Hz) gamma rays from the
blazar Mkn 501 (Aharonian et al 1999).  This is an
unexpected  result: the gamma
radiation should have severely been attentuated by photons from the
isotropic infra-red background which exists along
the intervening line-of-sight, because of
pair production in the center-of-mass
frame $\Sigma'$ where both photon populations appear to be $>$
1 MeV in energy and undergoing essentially head-on collisions.

Note however a caveat in the logic: while all the relevant data
are collected only aboard the $\Sigma$ frame the
interaction between 25 TeV and infra-red photons has never directly
been studied experimentally in any detail
from this frame.  Rather, we simply relied on ${\cal L}$
to avail ourselves a view of the situation  from frame $\Sigma'$
where the physics of the interaction becomes well established but
the input parameters for the two colliding photons assume
values as derived from the transformed data of $\Sigma$.  To
scrutinize the performance of ${\cal L}$, let us focus on
the $\omega \rightarrow \omega'$ operation.
Since the two frames are
connected by a Lorentz factor of $\gamma > 107$, subsitution of the
above value of $\omega$ and $\sigma_t \approx t_P$
into Eq (11) reveals that $\sigma_{\omega'}/\omega' > 0.2$,
i.e. intrinsic uncertainties in the transformed energy of the original
gamma rays have become significant.  Thus, with respect to $\Sigma'$,
those 1 MeV photons from the blazar may assume quite different energies
than our prediction based upon the laboratory data and ${\cal L}$.
Owing to the sensitivity of the pair production rate to input energies (
particularly the presence of
a low energy threshold, below which the reaction cannot proceed)
we can no longer insist that the process has 
necessarily played the expected role
in attenuating the source radiation.  In fact it is
obvious that when $\hbar \omega \sim $ 25 TeV the photon
can be  transmitted without undergoing pair production
(because its actual energy in frame $\Sigma'$ turns out
to be sub-threshold).  Moreover the probability of such an
occurence increases as
as $\hbar \omega$
exceeds 25 TeV.

In our second example attention is drawn to an expected
cutoff in the cosmic ray
(CR) spectrum at energies above $E = 10^{20}$ eV,
as when
this laboratory frame energy measurement is transformed along
with the cosmic microwave background measurements to $\Sigma'$, the rest
frame of the particle (presumed to be a proton), the familar scene
of photo-pion production becomes apparent (Greisen 1966).  In reality, 
no such cutoff was observed (e.g. Takeda et al 1998).  Note, however, that  
the relevant Lorentz factor here is $\gamma \sim 10^{11}$, and if we 
substitute such a value of $\gamma$
and the above value of $E$ into Eq (12) we will
arrive at $\sigma_{E'}/E'
\sim 10^{14}$.  Such an absurd uncertainty renders obsolete any theoretical
expectations of the fate of the particle which is based upon a
Lorentz transformation to the rest frame (another manner of expression is
to say that
because the rest frame cannot at all be identified the incident photon
energy in this frame is unknown).  
Certainly 50 \% of the particles
can avoid an interaction: for these particles the energy of the
microwave photon relative to $\Sigma'$ is below the pion
production threshold.
Moreover, the colossal value of $\sigma_{E'}$
prevents any modification of our conclusion
even when multiple in-situ interactions {\it and}
the improbable scenario of $\sigma_t < t_P$ are invoked.

For very high energy cosmic rays we wish to inquire the
value of the `critical energy' $E_c$ beyond
which the notion of
a unique rest frame is no longer tenable.  By means of 
Eq (12), one readily deduces that $E_c \approx 2 \times 10^{15}$ eV for
protons, and $\approx 3 \times 10^{13}$ eV for electrons.

An earlier work (Lieu 2001) presented a plausibility argument
for the existence of space and time units which vary
randomly both in magnitude and direction, by showing that such
a model can interpret relativity as the macroscopic
(aggregate) behavior of a microscopic ensemble.
The above discussion provides a direct means of testing (a) the
existence of Planck scale space-time fluctuations and (b) the
validity of the Principle of Relativity as applied to  ultra-relativistic
speeds.  If both hypotheses are upheld the reaction rates for the
two processes considered should exhibit an anomaly in the forementioned
situations.
The gamma ray and cosmic ray data gathered to date suggest
that the said anomaly is indeed in place.

It is also worth mentioning that the phenomenon discussed
has implications on the theory of General Relativity.
Since in the framework of this theory  gravity is a sequence of local
Lorentz frames $\Sigma'$ moving at varying speeds relative
to the laboratory frame $\Sigma$, our everyday perception of
physical laws are to be applied to $\Sigma'$ and then transformed
to $\Sigma$ - a transformation which again exaggerates the 
space-time uncertainties beyond their Planck scale values.
Effectively this means that space and time in a gravitation field
must fluctuate more severely than the levels given by
$l_P$ and $t_P$.  Obviously the consequences in the case of
strong fields are very interesting.

Author thanks W.I. Axford and Y. Takahashi for helpful discussions.

\noindent
{\bf References}

\noindent
Abramovici, A. {\it et al}, {\it Phys. Lett. A}, {\bf 218}, 157 (1996).\\
\noindent
Aharonian, F.A. {\it et al}, {\it Astron. Astrophys.}, {\bf 349},
11A (1999).\\
\noindent
Camelia, G.A., {\it Nature}, {\bf 410}, 1065 (2001). \\
\noindent
Greisen, K., {\it Phys. Rev. Lett.}, {\bf 16}, 748 (1966).\\
\noindent
Lieu, R., {\it Foundations of Physics}, {\bf 31}, 1233 (2001).\\
\noindent
Takeda, M. {\it et al}, {\it Phys. Rev. Lett.}, {\bf 81}, 1163 (1998).\\
\noindent
Wilczek, F., {\it Physics Today}, {\bf 54}, 13 (2001).\\

\end{document}